\documentclass[aps,showpacs,twocolumn,floatfix,bibnotes]{revtex4}

\usepackage{here}

\usepackage[dvips]{graphicx}

\usepackage[dvips]{graphicx}

\newcommand\pictc[5]{\begin{figure}
                       \centerline{
                       \includegraphics[width=#1\columnwidth, keepaspectratio, height=0.8\textheight]{#3}}
                   \protect\caption{\protect\label{fig:#4} #5}
                    \end{figure}            }
\newcommand\pict[4][1.]{\pictc{#1}{!tb}{#2}{#3}{#4}}
\newcommand\rpict[1]{\ref{fig:#1}}

\newcommand\leqt[1]{\protect\label{eq:#1}}
\newcommand\reqtn[1]{\ref{eq:#1}}
\newcommand\reqt[1]{(\reqtn{#1})}

\newcounter{Fig}

\begin{document}
\begin{sloppy}

\title{Spatial solitons in nonlinear left-handed metamaterials}

\author{Ilya V. Shadrivov}
\author{Yuri S. Kivshar}
\affiliation{Nonlinear Physics Group, Research School of Physical
Sciences and Engineering, Australian National University,
Canberra ACT 0200, Australia}

\begin{abstract}
We predict that nonlinear left-handed metamaterials can support
both TE- and TM-polarized self-trapped localized beams, {\em
spatial electromagnetic  solitons}. Such solitons appear as
single- and multi-hump beams, being either symmetric or
antisymmetric, and they can exist due to the hysteresis-type
magnetic nonlinearity and the effective domains of negative
magnetic permeability.
\end{abstract}

\pacs{78.20.Ci, 41.20.Jb, 42.81.Dp}

\maketitle

Left-handed metamaterials, i.e. materials with simultaneously
negative real parts of dielectric permittivity and magnetic
permeability~\cite{Veselago:1967-517:UFN}, demonstrate many
peculiar properties including negative refraction and
subwavelength imaging. A number of recent experiments reported the
observation of some of those properties for artificially
fabricated composites~\cite{science,boing} and planar
transmission-line structures~\cite{elef} at microwave frequencies.
There are strong efforts to fabricate the left-handed composite
materials operating at other frequencies, and negative magnetic
permeability at THz frequencies has been recently demonstrated in
planar structures composed of nonmagnetic conductive resonant
elements~\cite{science2}.

Composite materials used for demonstrating the properties of
left-handed media and negative refraction possess combined
negative electric and magnetic response in some {\em finite
frequency range} determined by the resonant properties of
nonmagnetic conductive elements of the artificial composites.
Nonlinearity introduced in such materials would allow to control
the frequency domain with the left-handed properties through strong modification of the effective nonlinear magnetic response~\cite{Zharov:2003-37401:PRL}. In particular, both dielectric permittivity and magnetic permeability of the composite
material consisting of a mesh of wires and split-ring resonators
embedded into a nonlinear dielectric can be controlled by varying
the electromagnetic field intensity in such a way that it may
allow switching the material properties between left- and
right-handed~\cite{Zharov:2003-37401:PRL}. Since microscopic
fields can be enhanced dramatically in a composite structure due
to the split-ring resonators, strong nonlinear effects are
expected to be important in such types of metamaterials, and this
can stimulate applications of nonlinear left-handed materials for
switching devices, limiters, and frequency conversion. In
addition, different types of nonlinear elements can also be easily
introduced into planar left-handed transmission-line
structures~\cite{elef} making a response of such a structure
strongly nonlinear.

In this Letter, we study, for the first time to our knowledge,
self-trapping of electromagnetic waves and the formation of
spatial solitons in nonlinear left-handed composite media, and
find the conditions for the existence of both TE- and TM-polarized
bright and dark solitons~\cite{Kivshar:2003:OpticalSolitons}. We
demonstrate that left-handed materials with a hysteresis-type
(multi-stable) nonlinear magnetic response support novel and
unique types of single- and multi-hump (symmetric, antisymmetric,
or even asymmetric) backward-wave spatial electromagnetic solitons
due to the effective domains corresponding to different branches
of nonlinear magnetic permeability.

We consider a nonlinear composite structure described, within an
effective medium approximation, by the nonlinear magnetic
permeability of the form~\cite{Zharov:2003-37401:PRL}
\begin{equation} \leqt{permeability_1}
\mu(\omega; |H|^2) = 1 + \frac{F \omega^2}{\omega_{0NL}^2 (|H|^2)
- \omega^2 +i\Gamma\omega},
\end{equation}
where $H$ is the applied magnetic field, $F$ is a filling factor
of the composite, $\omega_{0NL}$ is a nonlinear eigenfrequency of
the split-ring resonator,  and $\Gamma$ is the loss coefficient.
We assume that the infilling dielectric of the structure has a
Kerr-like nonlinear response,
\begin{equation} \leqt{kerr_epsilon}
\epsilon_{D}\left( |E|^2 \right) = \epsilon_{D0} + \alpha |E|^2/E_c^2,
\end{equation}
where $\epsilon_{D0}$ is the linear dielectric permittivity, $E_c$
is a characteristic electric field of a nonlinear response, and
$\alpha = \pm 1$ stands for focussing or defocusing nonlinearity,
respectively. Here we neglect the dependence of the effective
dielectric permittivity on the electric field assuming that a
nonlinear dielectric fills only the slits of the split-ring
resonators, as discussed in Ref.~\cite{Zharov:2003-37401:PRL}.

Dependence of the nonlinear eigenfrequency of split-ring
resonators on the {\em macroscopic} magnetic field $H$ can be
found in an implicit form (see Eq.~(11) of Ref.
~\cite{Zharov:2003-37401:PRL}),
\begin{equation} \leqt{magn_field}
|H|^2 = \alpha A^2 \frac{ \left( 1-X^2 \right)\left[
                                \left(X^2-\Omega^2\right)^2
                                +\Omega^2\gamma^2 \right]}{X^6},
\end{equation}
where $X = \omega_{0NL}/\omega_{0}$ is the normalized nonlinear
frequency, $A^2=16\epsilon_{D0}^3 \omega_{0}^2h^2/c^2$,
$\Omega=\omega/\omega_{0}$, $\omega_{0}=(c/a)[d_g / \pi h
\epsilon_{D0}]^{1/2}$ is the linear frequency of a split-ring
resonator, $\Gamma^{\prime} = \Gamma/\omega_{0}$ is the normalized
damping coefficient, $c$ is the speed of light, $d_g$, $h$, $a$
are the resonator parameters, and the magnetic field $H$ is
normalized to the critical electric field $E_c$.

When the losses are negligible, the dependence of the magnetic
permeability on the normalized magnetic field $H$ can have several
distinct forms~\cite{Zharov:2003-37401:PRL}. The most interesting,
multi-valued dependence is shown in the inset
Fig.~\rpict{phase_diagram} for the defocusing nonlinearity and
$\Omega > 1$.

\pict{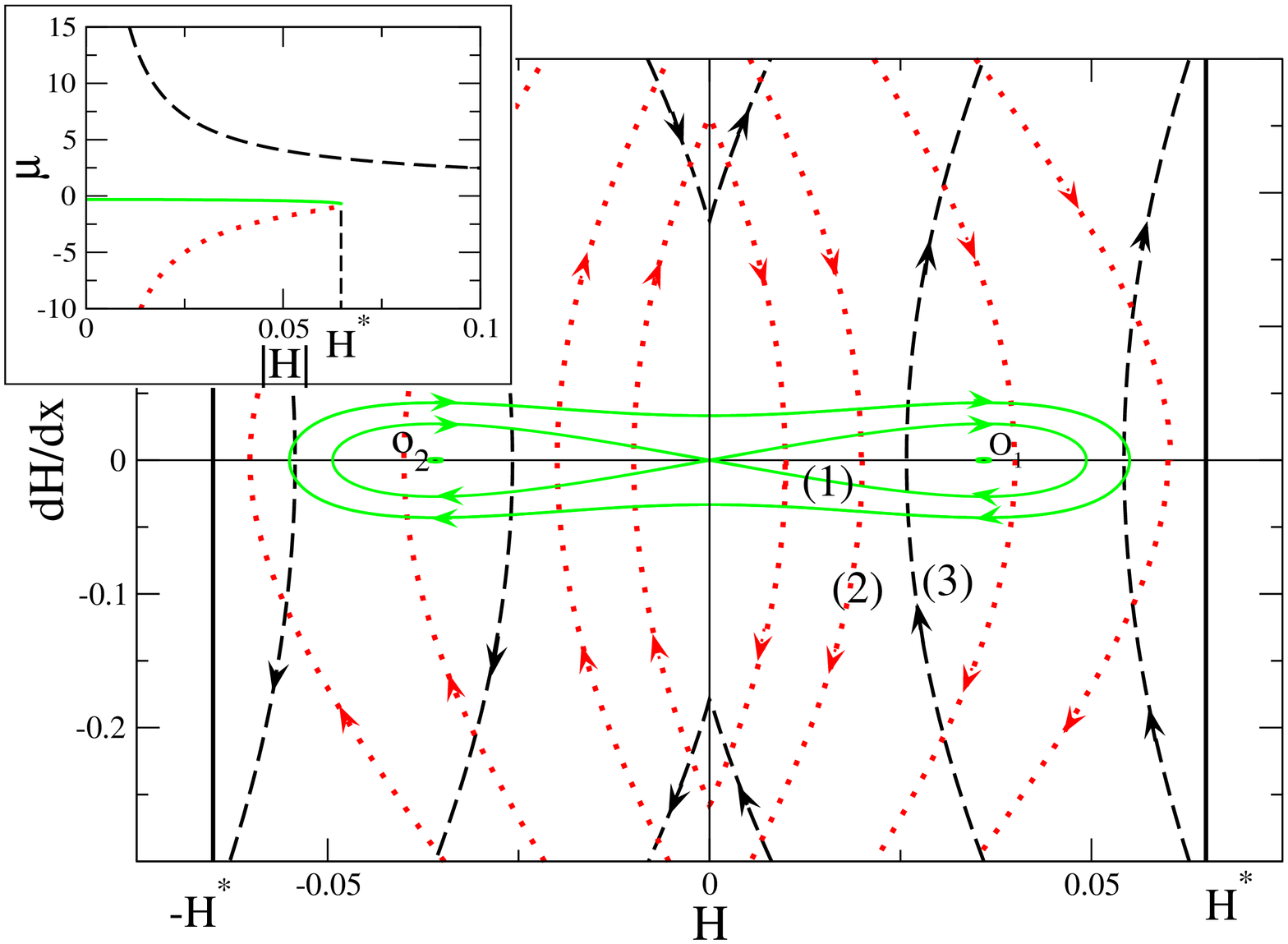}{phase_diagram}{(color online) Phase
trajectories on the plane $(H, dH/dx)$ described by
Eq.~\reqt{H_field} at $\gamma = 2.8$ with the nonlinear magnetic
response shown in the inset for $\alpha = -1$, $\epsilon_{D0} =
12.8$, $h = 0.03$ cm, $d_g = 0.01$ cm, $a = 0.3$ cm, and $F =
0.4$. Solid, dashed, and dotted lines mark the phase trajectories
corresponding to the solid, dashed and dotted branches of the
nonlinear magnetic permeability.}

{\em TM-polarized solitons.} First, we study spatially localized
TM-polarized waves that are described by one component of the
magnetic field and two components of the electric field. We
consider monochromatic stationary waves  with the magnetic field
component $H = H_y$ propagating along the $z$-axis and homogeneous
in the $y$-direction, $[\sim \exp{(i\omega t - i k z)}]$,
described by the dimensionless nonlinear Helmholtz equation
\begin{equation}\leqt{H_field}
\frac{d^2 H}{d x^2} + [\epsilon \mu(|H|^2) - \gamma^2]H = 0,
\end{equation}
where $\gamma = k c/\omega$ is a wavenumber, $x =
x^{\prime}\omega/c$ is the dimensionless coordinate, and
$x^{\prime}$ is the dimensional coordinate. Different types of
localized solutions of Eq.~\reqt{H_field} can be analyzed on the
phase plane $(H, dH/dx)$ (see, e.g.,
Refs.~\cite{Gildenburg:1983-48:ZETF,Rabinovich:1989:Oscillations}).
First, we find the equilibrium points: the point $(0,0)$ existing
for all parameters, and the point $(0,H_1)$, where $H_1$ is found
from Eq.~\reqt{magn_field} at
\begin{equation}\leqt{X_eq}
X^2(H_1) = X^2_{\rm eq} = \Omega^2 \left\{1 + \frac{F\epsilon_{\rm
eff}} {(\gamma^2 - \epsilon_{\rm eff})} \right\}.
\end{equation}
Below the threshold, i.e. for $\gamma < \gamma_{\rm tr}$, where
$\gamma_{\rm tr}^2 = \epsilon [1+F\Omega^2/(1-\Omega^2)]$, the
only equilibrium state $(0,0)$ is a saddle point and, therefore,
no finite-amplitude or localized waves can exist. Above the
threshold value, i.e. for $\gamma > \gamma_{\rm tr}$, the phase
plane has three equilibrium points, as shown in
Fig.~\rpict{phase_diagram} for the particular value $\gamma =
2.8$.

\pict{fig02.eps}{solitons}{(color online) Examples of different types of solitons:
(a) fundamental soliton; (b,c) solitons with one domain of
negative or positive magnetic permeability (shaded), respectively;
(d) soliton with two different domains (shaded). Insets in (b,c)
show the magnified regions of the steep change of the magnetic
field.}

In the vicinity of the equilibrium state $(0,0)$, linear solutions
of Eq.~\reqt{H_field} describe either exponentially growing or
exponentially decaying modes. The equilibrium state $(0, H_1)$
describes a finite-amplitude wave mode of the transverse
electromagnetic field. In the region of multi-stability, the type
of the phase trajectories is defined by the corresponding branch
of the multi-valued magnetic permeability. Correspondingly,
different types of the spatial solitons appear when the phase
trajectories correspond to the different branches of the nonlinear
magnetic permeability.

 The fundamental soliton is described by the
separatrix trajectory on the plane $(H, dH/dx)$ that starts at the
point $(0, 0)$, goes around the center point $(0, H_1)$, and then
returns back [see Fig.~\rpict{phase_diagram}, solid curve]; the
corresponding soliton profile is shown in
Fig.~\rpict{solitons}(a). More complex solitons are formed when
the magnetic permeability becomes multi-valued and is described by
several branches. Then, soliton solutions are obtained by
switching between the separatrix trajectories corresponding to
different (upper and lower) branches of magnetic permeability.
Continuity of the tangential components of the electric and
magnetic fields at the boundaries of the domains with different
values of magnetic permeability implies that both $H$ and $dH/dx$
should be continuous. As a result, the transitions between
different phase trajectories should be continuous.

Figures~\rpict{solitons}(b,c) show several examples of the more
complex solitons corresponding to a single jump to the lower
branch of $\mu(H)$ (dotted) and to the upper branch of $\mu(H)$
(dashed), respectively. The insets show the magnified domains of a
steep change of the magnetic field. Both the magnetic field and
its derivative, proportional to the tangential component of the
electric field, are continuous. The shaded areas show the
effective domains where the value of magnetic permeability
changes. Figure~\rpict{solitons}(d) shows an example of more
complicated multi-hump soliton which includes two domains of the
effective magnetic permeability, one described by the lower
branch, and the other one -- by the upper branch. In a similar
way, we can find more complicated solitons with different number
of domains of the effective magnetic permeability.

We note that some of the phase trajectories have discontinuity of
the derivative at $H=0$ caused by infinite values of the magnetic
permeability at the corresponding branch of $\mu_{\rm eff}(H)$.
Such a non-physical effect is an artifact of the lossless model of
a left-handed nonlinear composite considered here for the analysis
of the soliton solutions. In more realistic models that include
losses, the region of multi-stability does not extend to the point
$H = 0$, and in this limit the magnetic permeability remains a
single-valued function of the magnetic
field~\cite{Zharov:2003-37401:PRL}.

\pict{fig03.eps}{nl_disp}{(color online) Normalized soliton energy flow vs. its
propagation constant. Vertical solid line is a boundary of wave
localization in the linear limit, $\gamma = \gamma_{\rm tr}$. Thin vertical dashed
line shows the existence boundary for the fundamental soliton.
Thick solid curve shows the dispersion of a fundamental soliton.
Dashed and dotted curves correspond to the solitons with one
domain where the value of magnetic permeability changes. Dotted
curves correspond to one-hump solitons shown in
Fig.~\rpict{solitons}(b), dashed curves correspond to the two-hump
solitons shown in Fig.~\rpict{solitons}(c).}

For such a multi-valued nonlinear magnetic response, the domains
with different values of the magnetic permeability "excited" by
the spatial soliton can be viewed as effective induced left-handed
waveguides which make possible the existence of single- and
multi-hump soliton structures. Due to the existence of such
domains, the solitons can be not only symmetric, but also
antisymmetric and even asymmetric. Formally, the size of an
effective domain can be much smaller than the wavelength and,
therefore, there exists an applicability limit for the obtained
results to describe nonlinear waves in realistic composite
structures.

Nonlinear dispersion of several types of spatial solitons is shown
in Fig.~\rpict{nl_disp} as the dependence of the soliton energy
flow vs. the normalized propagation constant $\gamma$. The
fundamental soliton exists in a limited range of $\gamma$. The
lower boundary is determined by the threshold value $\gamma_{\rm
tr}$. At the upper boundary, the amplitude of the soliton field
exceeds the critical value $H^*$, and the magnetic permeability
becomes positive. All spatial solitons are backward propagating
waves here, since the energy propagates in the direction opposite
to the wavevector, as indicated by the negative sign of power
flow.

\pict{fig04.eps}{phase_diagram_4}{(color online) Phase diagram of
Eq.~\reqt{HxHz} corresponding to the multi-valued magnetic
response shown in Fig.~\rpict{phase_diagram}(inset). Solid, dashed
and dotted curves mark the phase trajectories that correspond to
solid, dashed and dotted curves in
Fig.~\rpict{phase_diagram}(inset), respectively. Inside the circle
with the radius $H^*$ the magnetic permeability is a three-valued
function of $H$, but it is single-valued, otherwise. Thick solid
figure-eight curve is defined by a singularity in \reqt{HxHz}, at
which the (dotted) phase trajectories change their direction.}

When the infilling dielectric of the structure displays {\em
self-focusing nonlinear response}, we have $\Omega <1$, and in
such system we can find {\em dark solitons}, i.e. localized dips
on the finite-amplitude background
wave~\cite{Kivshar:2003:OpticalSolitons}.  Similar to bright
solitons, there exist both fundamental dark solitons and dark
solitons with domains of different values of magnetic
permeability. For self-defocusing nonlinearity and $\Omega<1$,
magnetic permeability is a single-valued function, and such a
nonlinear response can support dark solitons as well, whereas for
self-focusing dielectric, and $\Omega> 1$, no solitons
can exist. Thus, in the composite with self-defocusing dielectric, depending on the frequency of the electromagnetic field, either bright or dark solitons can exist.

{\em TE-polarized solitons.} The TE-polarized waves are described
by one component of the electric field and two components of the
magnetic field. For such waves, magnetic permeability  depends on
two components of the magnetic field and the analysis of localized
modes becomes more involved.

We look for the waves propagating in the $z-$direction,
\[
{\bf E} = {\bf y_0} E e^{i\omega t - i h z}, \;\;\; {\bf H} =
({\bf x_0} H_x + {\bf z_0} iH_z ) e^{i\omega t - i h z},
\]
where $({\bf x_0}, {\bf y_0}, {\bf z_0})$ are the unit vectors,
and $E$, $H_x$, and $H_z$ are scalar real amplitudes of the
fields. A constant phase difference between the components $E$ and $H_z$ is
assumed, and no energy flow in the $x$-direction occurs. Spatial distribution of the magnetic field in a soliton is described by a system of coupled equations,
\begin{eqnarray}\leqt{HxHz}
\frac{d H_z}{dx} = [\epsilon \mu\left(H^2\right) -
\gamma^2]H_x/\gamma,\nonumber\\
\frac{d H_x}{dx} = -\frac{\gamma^2\mu\left(H^2\right) +
                            \xi \left[\epsilon \mu\left(H^2\right) -
                            \gamma^2 \right]}{
                            [\gamma \mu\left(H^2\right) + \gamma\xi]},
\end{eqnarray}
where $\xi = 2 H_x^2 \mu^{\prime}(H^2)$, $H^2 = H_x^2+H_z^2$. We
analyze Eqs.~\reqt{HxHz} on the plane $(H_x, H_z)$, as shown in
Fig.~\rpict{phase_diagram_4}.

Similar to the TM-polarized waves, the phase trajectory on the
plane $(H_x, H_z)$ is defined by a particular branch of the
multi-valued nonlinear magnetic permeability. Near the origin $(0,
0)$, the phase trajectory is defined by the solid branch of the
magnetic permeability [see Fig.~\rpict{phase_diagram}(inset)]. In
general, the phase diagram has three equilibrium states, which are
similar to those described above for the TM-polarized waves. A
singularity curve (thick solid line) is defined by the vanishing
denominator in the second equation of Eqs.~\reqt{HxHz}. The phase
trajectories represented by dotted curves change their direction
when they cross the singularity line.

\pict{fig05.eps}{nl_disp2}{(color online) Energy flow of the fundamental soliton
(solid) and two solitons with a single domain (dashed and dotted).
Solitons characterized by the dashed curves (left inset) have a
single domain (shaded) with magnetic permeability described by the
dashed curve in Fig.~\rpict{phase_diagram}(inset). Solitons
characterized by the dotted curves (right inset) have a domain
(shaded) with magnetic permeability described by the dotted curve
in Fig.~\rpict{phase_diagram}(a, inset). Examples show the
profiles of the components $H_x$ (solid) and $H_z$ (dashed) in the
TE soliton.}

For a homogeneous structure  with the nonlinear magnetic
permeability shown by a solid curve in
Fig.~\rpict{phase_diagram}(inset), the separatrix curve on the
phase plane of Fig.~\rpict{phase_diagram_4} describes the
fundamental two-component TE-polarized spatial solitons with a
smooth envelope. However, there exist more complex TE-polarized
solitons, and such solitons include effective domains with
different values of the multi-valued magnetic permeability. At the
boundaries of the domains, both the fields $H_x$ and $H_z/\mu$
should be continuous. As a result, the phase trajectories
describing the TE-polarized solitons with such domains will be
discontinuous on the plane $(H_x, H_z)$. In Fig.~\rpict{nl_disp2},
we show the energy flow for the family of the fundamental solitons
and two types of solitons with a single domain, as the functions
of the normalized propagation constant $\gamma$.  Example of the
solitons presented here correspond to a symmetric dependence of
the $x$-component of the magnetic field; however, both
antisymmetric and asymmetric TE-polarized solitons can exist in
the structure as well.

In conclusion, we have analyzed both TE- and TM-polarized
self-trapped nonlinear localized beams -- spatial electromagnetic
solitons -- in left-handed metamaterials with nonlinear resonant
response. We have revealed the existence of unique types of
solitons supported by the hysteresis-type magnetic nonlinearity
and the domains with different values of magnetic permeability.
Such solitons can have symmetric, antisymmetric, or asymmetric
structure. We believe that similar solitons can be found in other
types of complex nonlinear materials with left-handed properties
or the frequency-dependent domains corresponding to negative
refraction.

We thank Alex Zharov for useful discussions and acknowledge a
support of the Australian Research Council.

\end{sloppy}
\end{document}